# Theoretical and experimental study of photoacoustic excitation of silica-coated gold nanospheres in water

*Genny A. Pang[1a], Florian Poisson[2], Jan Laufer[3], Christoph Haisch[1], Emmanuel Bossy[2b]*

[1] Chair for Analytical Chemistry and Institute of Hydrochemistry, Technische Universität München, Marchioninistraße 17, 81377, Munich, Germany

[2] Univ. Grenoble Alpes, CNRS, LIPhy, 38000 Grenoble, France

[3] Institut für Physik, Martin-Luther-Universität Halle-Wittenberg, Halle (Saale), Germany

[a] E-Mail: genny.pang@tum.de

[b] E-Mail: emmanuel.bossy@univ-grenoble-alpes.fr

**Abstract**

Silica-coated gold nanoparticles are commonly employed in biomedical photoacoustic (PA) imaging applications. We investigate theoretically and experimentally the PA signal generation by silica-coated gold nanospheres in water. Our theoretical model considers thermoelastic expansion in the long-pulse illumination regime, and the PA signals are determined based on a semi-analytical solution to the thermal diffusion equations and a finite-difference in time domain (FDTD) solution to the thermoelastic equations. Both the influence of interfacial thermal (Kapitza) resistance at the gold-water boundary and the influence of the silica coating on PA signal generation were investigated. Our results indicate that for the nanosecond pulses commonly employed in PA imaging, Kapitza resistance has a negligible effect on photoacoustic signal generation. Moreover, our model shows that the presence of a silica coating causes a reduction in the PA signal amplitude, with the level of signal reduction increasing with thicker silica coating. Our theoretical predictions are qualitatively consistent with our experimental results, where suspensions of in-house-synthesized and commercially available silica-coated gold nanosphere suspensions were excited with nanosecond-pulsed laser illumination at 532 nm. The PA signal amplitudes from silica-coated nanospheres were lower than the signal amplitudes for uncoated gold nanospheres of the same core gold diameter. The amount of reduction of the experimentally PA signal amplitude due to the silica coating was found to increase with thicker silica coating, in agreement with our theoretical predictions.

**Keywords:** photoacoustic imaging, plasmonic nanoparticle, gold nanosphere, silica coating, Kapitza resistance, photothermal effect

Silica coating of gold nanoparticles for biomedical applications provides several advantages, including enhanced colloidal stability,[1-4] increased cellular uptake[5], low toxicity,[1, 6, 7] increased thermal stability and resistance to melting,[7-10] and increased surface area and porosity improving functionalization possibilities with targeting moieties.[1, 2, 6, 11] Imaging of these silica-coated gold nanoparticles can be advantageous to provide information about the particle uptake, distribution, and clearance. Photoacoustic (PA) imaging is ideal for biomedical applications due to its deep imaging penetration depth, capability of spectroscopic contrast, enabling visualization of nanoparticles at depths up to several centimeters under the tissue surface.[12] PA imaging has already been applied to several biomedical studies employing silica-coated gold nanoparticles.[5-7, 13, 14] Some studies show an enhancement effect of the PA signal due to the presence of a silica coating,[5, 15, 16] even though it has been noted that adding a coating of finite heat conductance and capacity could broaden a heat pulse and deteriorate the PA signal.[15] No comprehensive theoretical model has yet been developed to explain this phenomenon, and the effect of the silica coating on the fundamental PA signal generation process is still not theoretically well understood.

PA imaging is a noninvasive biomedical imaging modality based on the conversion of absorbed light energy to ultrasound energy by means of the thermoelastic effect, where the image contrast depends on the spatial distribution of optical absorption properties of the imaged medium. Therefore, highly absorbing nanoparticles with a gold core exhibit excellent contrast in PA images.[17-19] During PA excitation of nanoparticle colloidal suspensions, intensity-modulated light energy is absorbed by the nanoparticles, causing an increase in local particle temperature. The energy absorbed by the particle is rapidly transferred to the surrounding medium where the majority of the PA signal generation from thermoelastic expansion occurs.[16, 20, 21] Because this heat transfer process occurs on the order of nanoseconds for nanoparticles, which is typically the same time scale of the PA excitation for most practical implementation, thermal confinement cannot be assumed. Moreover, the PA signal generation process can be influenced by the interfacial thermal (Kapitza) resistance between the particle and surrounding fluid.[15, 16] Consequently, a simple analytical solution for the PA pressure generated by a colloidal nanoparticle suspension is not available. While it is accepted that the coating of nanoparticles with silica influences the thermal resistance and heat transfer rate between the particle and the suspension fluid, and thus also the strength of the PA signal,[22] a comprehensive model describing this effect for silica-coated gold nanoparticles still needs to be developed. While detailed numerical approaches to predict the PA pressure from a finite-size homogeneous solid sphere have been developed,[16, 21, 23] these have not been extended to examine the influence of a particle coating on the PA response. A recent finite element modeling study has employed commercial software for a basic investigation of PA signal from a gold-silica core-shell nanospheres in water,[24] however, insufficient description of the simulation parameters (namely the assumed Kapitza resistance) make it difficult to evaluate the significance of the reported results.

In this study, we investigate through complementary theoretical and experimental approaches the influence on PA generation of the silica coating of an individual gold nanosphere suspended in water. Our theoretical modeling accounts for energy conservation in the gold nanosphere and transient heat diffusion through the silica coating and water. Through this model, we examine both uncoated gold nanospheres and silica-coated gold nanospheres in water, and investigate the influence of Kapitza resistance at the particle-water boundary on the spatially- and temporally-resolved temperature profile, in terms of the resulting effects on the PA signal amplitude. The experimental component of our work consists of a systematic study of PA excitation of aqueous suspensions of uncoated gold nanospheres and silica-coated gold nanospheres with varying silica coating thicknesses from 0 to 25 nm. Both our theoretical and experimental results show that the presence of a silica coating on gold nanospheres causes a reduction in the PA signal.

## Results

**Theoretical Influence of Kapitza Resistance**

The effect of interfacial thermal resistance on the photoacoustic wave generation from pulsed-laser excitation of a single uncoated gold nanosphere in water with a finite Kapitza resistance at the gold-water interface, depicted in Figure 1a, was solved using a semi-analytical method combined with a finite-difference time-domain solution (FDTD). These results serve as a baseline to compare subsequent model calculations that examine the effect of silica coating. Our approach follows that used by Prost *et al.*,[21] who initially solved the problem without Kapitza resistance. In brief, the temperature field, governed by conservation of energy and thermal diffusion as given by Equations (1) and (2), respectively, are solved by using Laplace Transform with the boundary conditions given by Equations (3) and (4) to compute the impulse response of the governing equations. Integral expressions of the impulse temperature response are computed numerically for discrete values of time and position as required by the subsequent FDTD computations of the pressure field. The temperature field as a function of space and time in response to pulsed PA excitation with a finite pulse width is then determined by numerical convolution of the impulse response with the excitation source function, which we assumed as a Gaussian pulse. The temperature field solution is used as a source term in the thermo-elastic Equations (5), (6), and (7), the PA pressure, equal in water to the radial stress tensor $\sigma_{rr}$, is computed using a FDTD algorithm. Eq (5), (6) and (7) take into account thermal expansion in all the materials considered in the model. Spherical symmetry was assumed, with the center of the gold nanosphere as the center of symmetry. The temperature in the gold nanosphere is taken to be uniform because thermal relaxation within the gold particle occurs on the order of picoseconds.[23, 25] Table 1 presents the governing equations in the model, Table 2 lists the variables used in the governing equations, and Table 3 shows the physical properties assumed for the all the materials considered in the model.[26] The thermal expansion coefficient of water was taken as independent of temperature, as it was verified that for the excitation fluence considered in this work the temperature elevation were not high enough to yield non-linear behavior such as described in previous studies.[21, 25, 27] The absorption coefficient of the gold nanosphere were derived from Mie theory with optical constants from Johnson and Christy.[28] The full details of the model, including the impulse response solution, discretization of the thermoelastic equations, grid criteria for conversion, and solution procedures are provided in the Supporting Information.

Table 1. Governing equations.

| | | | | |
|---|---|---|---|---|
| Conservation of energy | $\frac{4}{3}\pi a^3 \rho_g c_{p1} \frac{\partial T_g}{\partial t} = 4\pi a^2 k_i \frac{\partial(T_i)}{\partial r}\big|_{r=a} + \sigma_a \Phi f(t)$, uncoated: $i=w$, else $i=s$ | | with coating | (1) |
| Diffusion equation | $\frac{1}{\alpha_w}\frac{\partial(rT_w)}{\partial t} = \frac{\partial^2(rT_w)}{\partial r^2}$ uncoated, | $\frac{1}{\alpha_s}\frac{\partial(rT_s)}{\partial t} = \frac{\partial^2(rT_s)}{\partial r^2}$ $\frac{1}{\alpha_w}\frac{\partial(rT_w)}{\partial t} = \frac{\partial^2(rT_w)}{\partial r^2}$ | with coating | (2) |
| Heat flow boundary conditions | $k_g \frac{\partial(T_g)}{\partial r}\big|_{r=a} = k_w \frac{\partial(T_w)}{\partial r}\big|_{r=a}$ uncoated, | $k_g \frac{\partial(T_g)}{\partial r}\big|_{r=a} = k_s \frac{\partial(T_s)}{\partial r}\big|_{r=a}$ $k_s \frac{\partial(T_s)}{\partial r}\big|_{r=a+e_{Si}} = k_w \frac{\partial(T_w)}{\partial r}\big|_{r=a+e_{Si}}$ | with coating | (3) |
| Temperature boundary conditions | $T_g(r=a) - T_w(r=a) = -k_w R_{thw-g}\frac{\partial(T_w)}{\partial r}\big|_{r=a}$ uncoated, | $T_g(r=a) = T_s(r=a)$ $T_s(r=a+e_{Si}) = T_w(r=a+e_{Si})$ | with coating | (4) |
| Radial stress equation | $\frac{\partial \sigma_{rr}}{\partial t}(r,t) = \left[\lambda\left(\frac{\partial}{\partial r}+\frac{2}{r}\right)+2\mu\frac{\partial}{\partial r}\right]v_r(r,t) - \left(\lambda+\frac{2}{3}\mu\right)\beta\frac{\partial T}{\partial t}(r,t)$ | | | (5) |
| Orthoradial stress equation | $\frac{\partial \sigma_{\theta\theta}}{\partial t}(r,t) = \left[\lambda\left(\frac{\partial}{\partial r}+\frac{2}{r}\right)+\frac{2\mu}{r}\right]v_r(r,t) - \left(\lambda(r)+\frac{2}{3}\mu(r)\right)\beta\frac{\partial T}{\partial t}(r,t)$ | | | (6) |
| Velocity equation | $\frac{\partial v_r}{\partial t}(r,t) = \frac{1}{\rho}\frac{\partial \sigma_{rr}}{\partial r}(r,t) + \frac{2}{r}(\sigma_{rr}(r,t) - \sigma_{\theta\theta}(r,t))$ | | | (7) |

Table 2. Notations in the governing equations.

| Variable | Description |
|---|---|
| $i$ | Subscript indicating the medium: $g$, $s$, and $w$ for gold, silica, and water, respectively |
| $a$ | Gold nanosphere radius |
| $c_s$ | Speed of sound |
| $c_p$ | Specific heat capacity |
| $e_{Si}$ | Silica coating thickness |
| $f$ | Heating source term |
| $k$ | Thermal conductivity |
| $R_{th_{w-g}}$ | Kapitza resistance at gold-water interface |
| $r$ | radial distance |
| $T$ | Temperature |
| $t$ | Time |
| $v_r$ | Radial displacement |
| $\alpha$ | Thermal diffusivity |
| $\beta$ | Thermal expansion coefficient |
| $\Phi$ | Excitation fluence |
| $\lambda$ | 1st Lamé coefficient |
| $\mu$ | 2nd Lamé coefficient |
| $\rho$ | Density |
| $\sigma_a$ | Absorption cross section of gold nanosphere |
| $\sigma_{rr}$ | Radial stress tensor (PA pressure) |
| $\sigma_{\theta\theta}$ | Angular stress tensor |
| $\tau_p$ | Duration of excitation pulse |

Table 3. Properties of gold, silica and water at $T \approx 25$ °C.[26]

|  | Density $\rho$ (kg m$^{-3}$) | Specific heat $c_p$ (J K$^{-1}$ kg$^{-1}$) | Thermal conductivity $k$ (W m$^{-1}$ K$^{-1}$) | Thermal expansion coefficient $\beta$ (K$^{-1}$) | 1st Lamé coefficient $\lambda$ | 2nd Lamé coefficient $\mu$ | Speed of sound $c$ (m s$^{-1}$) |
|---|---|---|---|---|---|---|---|
| Gold | 19300 | 129 | 318 | 4.26x10$^{-5}$ | 147x10$^9$ | 27.8x10$^9$ | 3240 |
| Silica | 2200 | 740 | 1.38 | 0.55x10$^{-6}$ | 15.9x10$^9$ | 31.3x10$^9$ | 5973 |
| Water | 1000 | 4128 | 0.598 | 2.10x10$^{-4}$ | 2.25x10$^9$ | 0 | 1500 |

The temperature profiles of a PA excited uncoated gold nanosphere $a = 10$ nm with the Kapitza resistance $R_{th_{w-g}} = 1 \times 10^{-8}$ m$^2$K W$^{-1}$, as typical from the literature,[29] is shown in Figure 1 for a 5 ns pulse, a 1 ns pulse, and a 0.5 ns pulse. The temporal profiles of the particle core are shown in Figure 1b-d and the spatial temperature profiles at the peak temperature time showing the temperature distribution in water are shown in Figure 1e-g. For comparison, the simulated temperature solutions assuming no Kapitza resistance (solution of Prost et al.[21]) are also shown.

Our simulation results show that the expected finite $R_{th_{w-g}}$ leads to an increase in the peak temperature in the gold particle as compared to as if no Kapitza resistance was present, and a spatial discontinuity of temperature imposed by the model can be observed at the particle-water interface as shown in Figure 1e-g. The simulated PA signal for the same cases are shown in Figure 1h-j. For a pulse duration of 5 ns (Figure 1h), the change of temperature profile due to a finite $R_{th_{w-g}}$ has a negligible effect on the PA signal amplitude (< 0.5% influence on the PA signal), because the temperature profile in the surrounding water is negligibly influenced (as seen in Figure 1e). For shorter pulses, the PA signal amplitude is reduced by 4% for a pulse of 1 ns (Figure 1i) and 8% for a pulse of 0.5 ns (Figure 1j).

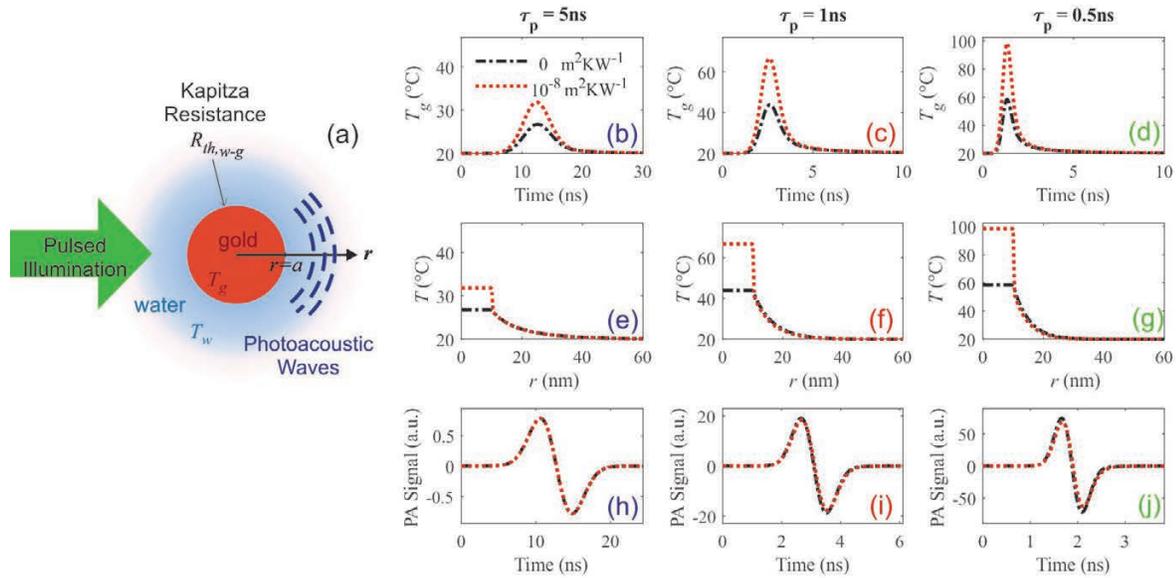

Figure 1. Simulation results of PA signal generation from a single uncoated gold nanosphere in water with finite Kapitza resistance at the gold-water interface shown in (a). (b)-(d) Temperature profile in time of the particle core, (e)-(g) temperature profile in space at peak temperature time, and (h)-(j) simulated PA signal measured 1 μm from particle center. Each result is for $a = 20$ nm after pulsed laser excitation $\Phi = 1$ mJ/cm² with no Kapitza resistance (black), and expected Kapitza resistance $R_{th_{w-g}} 1 \times 10^{-8}$ m²K W⁻¹ (red). Three different excitation pulse widths $\tau_p$ of 5, 1, and 0.5 ns are shown.

Figure 2a shows the sensitivity of the theoretical PA signal amplitude to the Kapitza resistance for $\tau_p = 5$, 1, and 0.5 ns for a gold nanosphere core of $a = 10$ nm. As a key output of this simulations, our results suggest that for pulse lengths of $\tau_p = 5$ ns, a typical pulse duration for PA experiments, the PA signal would only be expected to be influenced by Kapitza resistance for $R_{th_{g-w}} \geq 1 \times 10^{-7}$ m²K W⁻¹, which is an order of magnitude higher than the expected value at the gold-water interface of gold nanospheres in water. The result that Kapitza resistance at the gold-water interface does not influence the PA signal is consistent with our experimental results from a previous study.[25] When pulses shorter than 1 ns are considered, however, our simulations predicts that the PA signal is affected by the Kapitza resistance: the effect of $R_{th_{w-g}} = 1 \times 10^{-8}$ m²K W⁻¹ is to decrease the PA signal amplitude by 8% and 4% for a pulse duration of 0.1 ns and 1 ns, respectively. Figure 2b shows similar results for a larger particle core with $a = 20$ nm, illustrating similar trends, where the effect of $R_{th_{w-g}} = 1 \times 10^{-8}$ m²K W⁻¹ on the PA signal is negligible for $\tau_p = 5$ ns, however, reduces the PA signal when shorter excitation pulses are applied. Although a more detailed investigation for pulses shorter than 5 ns was out of the scope of this work, Figure 2 nevertheless clearly indicates that sub-nanosecond pulses allow probing the effect of Kapitza resistances, a result in agreement with previously reported time-resolved measurements of cooling dynamics showing sensitivity of femtosecond pulses to Kapitza resistances.[30]

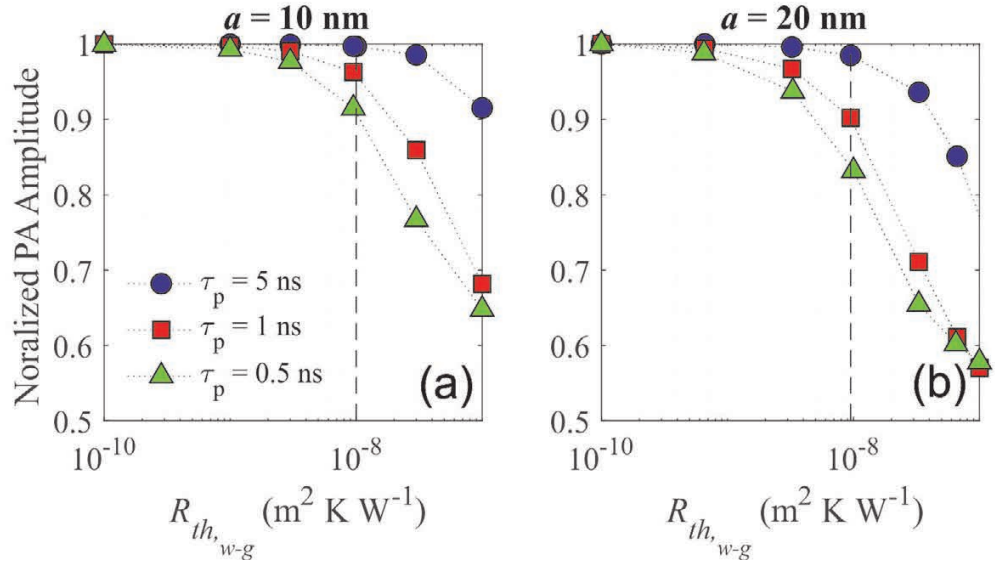

Figure 2. Simulation results PA signal amplitude normalized by value without Kapitza resistance as a function of $R_{th_{w-g}}$ for $\tau_p$ of 5, 1, and 0.5 ns for (a) $a = 10$ nm, and (b) $a = 20$ nm.

**Theoretical Influence of Silica Coating**

The effect of silica coating thickness ($e_{Si}$) on the photoacoustic wave generation from pulsed-laser excitation of a single silica-coated gold nanosphere in water, depicted in Figure 3a, was solved using a theoretical approach analogous to that described in the previous section. Equations (1) to (4) were modified as described in Table 1 to include the diffusion and heat flow through the silica coating. When applicable, the PA signal amplitude solved from the silica-coated gold nanosphere was normalized by the respective value from the uncoated gold nanosphere of identical core radius $a$. Because in this work, we are concerned only with PA excitation on the order of $\tau_p = 5$ ns, where the effect of Kapitza resistance is shown by Figure 2 to be negligible on the PA signal, the PA signals from the uncoated gold nanosphere were simulated with no Kapitza resistance. While the presence of a non-zero Kapitza resistance at the gold-silica or silica-water interface is possible, a lack of literature on the Kapitza resistance at the boundaries of a silica-coated gold nanosphere in water prevents a precise calculation of how the actual Kapitza resistance affects the PA signal. However, based on the available literature on Kapitza resistance in systems of similar types of coated metal nanospheres,[31] we do not expect the magnitude of such a Kapitza resistance to be greater than the order of magnitude of the Kapitza resistance at the gold-water interface. Therefore, for pulse excitation of $\tau_p = 5$ ns, Kapitza resistance at the boundaries of the silica-coated gold nanosphere would not be expected to have a significant effect in the PA signal, and thus will not be considered in our model.

Figure 3b-d present the effect of silica coating thickness, with $e_{Si}$ ranging from 0 to 40 nm, on the temperature field of a gold nanosphere ($a = 10$ nm) after PA excitation (Gaussian pulse FWHM $\tau_p = 5$ ns, $\Phi_0 = 1$ mJ/cm$^2$). The effect of different $e_{Si}$ on the simulated time-resolved temperature profile in the gold nanosphere core is shown in Figure 3b. The highest peak temperature in the gold nanosphere is reached with no silica coating ($e_{Si} = 0$), and increasing $e_{Si}$ leads to a decreasing peak temperature in the gold nanosphere. Figure 3c shows that the peak temperature at the particle-water interface also follows this trend, and is the highest without silica coating, and decreases for increasing $e_{Si}$. At a fixed point in water outside the particle, the temperature increases with increasing $e_{Si}$, with the peak temperature arriving at an earlier time for larger $e_{Si}$, as shown in Figure 3d. From a thermal perspective, the silica coating therefore acts to improve the heat transfer from the particle to its surrounding environment, as mentioned in earlier works.[15, 16] The spatial distribution of the temperature at the time of peak temperature for $e_{Si} = 20$ nm is shown in Figure 3e, showing that for a coated particle, the water temperature is highest at the particle-water interface at any given time, but the temperature at this interface is lower than that in the gold core. If the PA generation is assumed to take place mostly in water (as the thermal expansion coefficient in silica is about 400 times smaller than that of water), it should therefore be largely dictated by the temperature at the particle-water interface. Therefore, it can be expected that the PA generation from a gold nanosphere will be reduced by the addition of a silica coating, although the silica coating improves the heat transfer from gold to water.

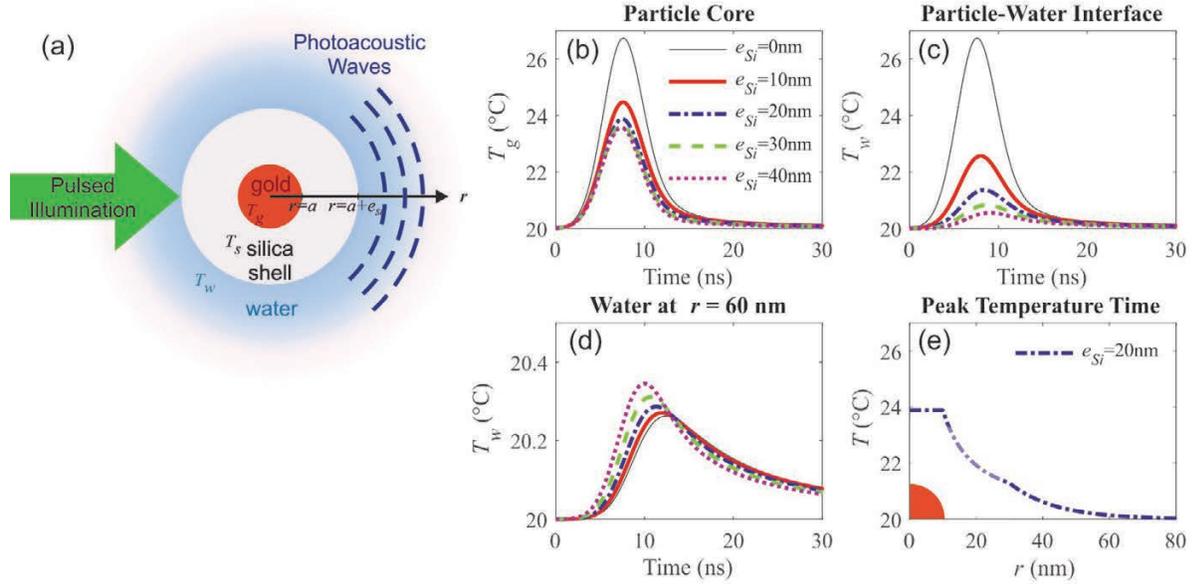

Figure 3. Solution of the temperature profile and PA signal from a single silica-coate gold nanosphere in water shown in (a). Time-varying temperature profile after excitation with $\tau_p = 5$ ns, $\Phi = 1$ mJ/cm² for $a = 10$ nm with $e_{Si}$ from 0 to 40 nm (b) within the gold nanosphere, (c) at the particle-water interface, and (d) at a fixed position in water 60 nm from the particle center. (d) Spatially-varying temperature profile at the time of the peak temperature for nanosphere with a $e_{Si} = 20$ nm.

The influence of the temperature profiles on the PA signal predicted by our model is shown in Figure 4a, which shows the PA signal from a gold nanosphere ($a = 10$ nm, $e_{Si} = 0$ to 40 nm) measured at a fixed distance in water from the nanosphere center ($r = 1$ μm). A larger $e_{Si}$ leads to a decrease in the PA signal amplitude, in agreement with our hypothesis considering the fact that the PA generation takes place in water.[15, 16, 21] The general trend of the PA signal amplitude reduction through increasing $e_{Si}$ is consistent for varying gold nanosphere sizes, as summarized in Figure 4b, which shows the PA signal amplitude as a function of $e_{Si}$, normalized by that with $e_{Si} = 0$ nm for the corresponding gold nanosphere core sizes of $a = 5, 10, 15,$ and $20$ nm.

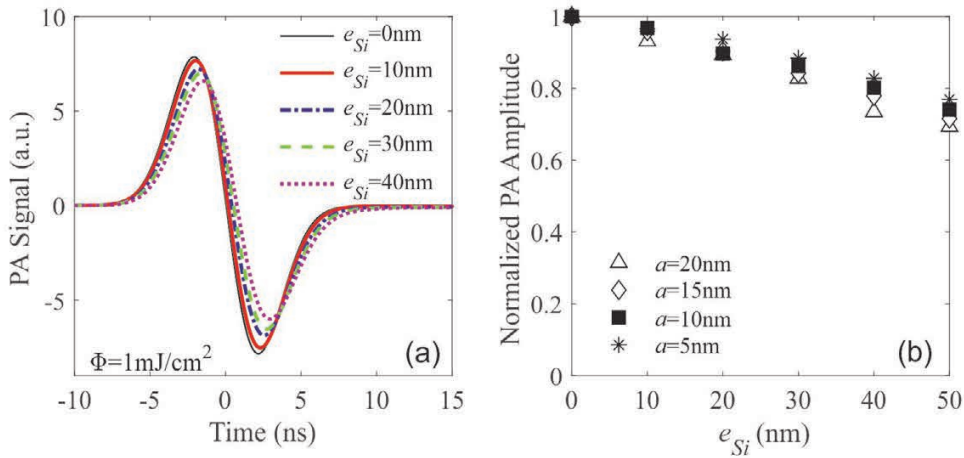

Figure 4. Simulated pulsed PA excitation with $\tau_p = 5$ ns, $\Phi = 1$ mJ/cm². (a) PA signal generated 1 μm from the nanosphere center for $a = 10$ nm with $e_{Si}$ between 0 and 40 nm. (b) PA signal amplitude with varying $e_{Si}$ for $a$ between 5 and 20 nm, normalized by the respective value with $e_{Si} = 0$ nm.

**Experimental Results**

Two different sources of aqueous suspensions of silica-coated gold nanospheres were experimentally studied. The first set of silica-coated gold nanospheres were synthesized in our laboratory, where citrate-stabilized gold nanospheres were primed with mPEG-thiol to enable transferability to isopropanol before a silica coating from tetraethyl orthosilicate (TEOS) was grown using a modified Stöber method.[32, 33] Two samples of silica-coated gold nanospheres were synthesized, one with $e_{Si} = 15$ nm and one with $e_{Si} = 25$ nm. Figure 5a-c presents scanning electron microscopy (SEM) images of the silica-coated and uncoated gold nanosphere suspensions, each with $a = 12.5$ nm. The optical extinction spectra of suspensions of these nanospheres diluted to an identical optical density of 0.75 at the wavelength of 532 nm are shown in Figure 5d. The second set of silica-coated gold nanospheres were obtained from a commercial source (Aldrich, $a = 5$ nm, $e_{Si} = 17$ nm). These were compared to uncoated gold nanospheres ($a = 5$ nm, $e_{Si} = 0$ nm), obtained from the same source. The SEM image and the optical extinction spectra of the commercially available nanosphere suspensions are provided in the Supporting Information (Figure S1).

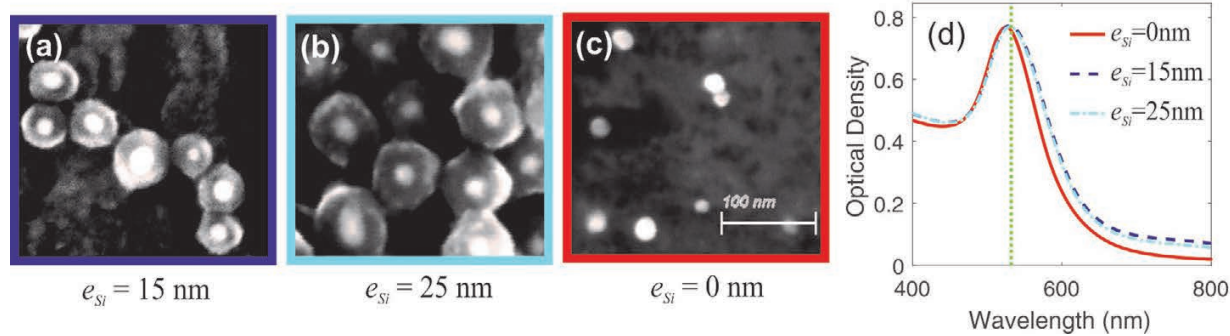

Figure 5. In-house-synthesized gold nanospheres $a = 12.5$ nm with and without silica coating. SEM images for (a) $e_{Si} = 15$ nm, (a) $e_{Si} = 25$ nm, and (a) $e_{Si} = 0$ nm; the scale bar of 100 nm is the same for all images. (d) OD spectra measured using a spectrometer in a 10 mm cuvette for suspensions all with an OD of 0.75 at the laser wavelength of 532 nm (shown by the dotted green line).

From measurements of the extinction and absorption spectra of the nanosphere suspensions in an integrating sphere setup, we found that the silica coating does not influence the fraction of the optical extinction due to scattering (see Supporting Information, Figure S2). Even for the silica-coated particles with an overall diameter of 75 nm ($a = 12.5$ nm, $e_{Si} = 25$ nm), the contribution of absorption to the optical extinction from the largest nanosphere is approximately 97% (Supporting Information, Figure S2c), which is in agreement with Mie theory predictions for a gold nanosphere in water with $a = 12.5$ nm.[34] Therefore, the optical density (OD), reported in this paper as that measured over a 10 mm optical path length, is assumed to represent the absorption characteristics of the particle suspensions.

Figure 6a shows representative PA signals at different dilutions of particle concentration (noted by their differing OD) for suspensions of the in-house-synthesized uncoated gold nanospheres ($a = 12.5$ nm) measured in a backwards-mode PA sensor with pulsed laser excitation $\tau_p = 5$ ns (FWHM) at 532 nm and excitation fluence ($\Phi$) of 3.5 mJ/cm$^2$. The first peak of the PA signal, shown at time zero, corresponds to the time that the acoustic signal from the illuminated sample surface reaches the acoustic detector. We define the PA signal amplitude as the magnitude of this first peak minus the background signal amplitude, which was observed with a repeatable fluence-dependent magnitude when a non-absorbing sample containing no nanospheres was placed on the PA sensor. As seen from Figure 6a, the suspension with lower OD exhibits a lower PA signal amplitude as expected, because fewer particles absorb less overall light energy. Figure 6b shows the comparison of the PA signal measured from the in-house-synthesized uncoated and silica-coated gold nanospheres ($a = 12.5$ nm, $e_{Si} = 0$ and 25 nm), showing that the PA signal amplitude from the suspension with silica coating is lower than that from the suspension of the uncoated nanospheres.

The relationship between OD and PA signal amplitude is summarized in Figure 6c, for the in-house-synthesized uncoated and silica-coated gold nanoparticle suspensions ($a = 12.5$ nm). Both the uncoated ($e_{Si} = 0$ nm) and coated particles ($e_{Si} = 25$ nm) show a linear increasing trend of PA signal amplitude with increasing OD, and the PA signal amplitude from the $e_{Si} = 25$ nm nanosphere suspension is consistently lower than that of the uncoated nanospheres. The error bars for all PA signal amplitudes shown in this work represent the standard deviation of the signal amplitude as measured in at least three measurement repetitions, with five consecutive signal traces captured for each measurement repetition. The low error bars of the measured PA signal amplitude with different suspension samples through several experiment repetitions led us to assume that particle aggregation was successfully avoided and the suspension was sufficiently disperse without non-uniform agglomerates that affect the PA signal. Figure 6d shows a schematic of the PA experimental setup.

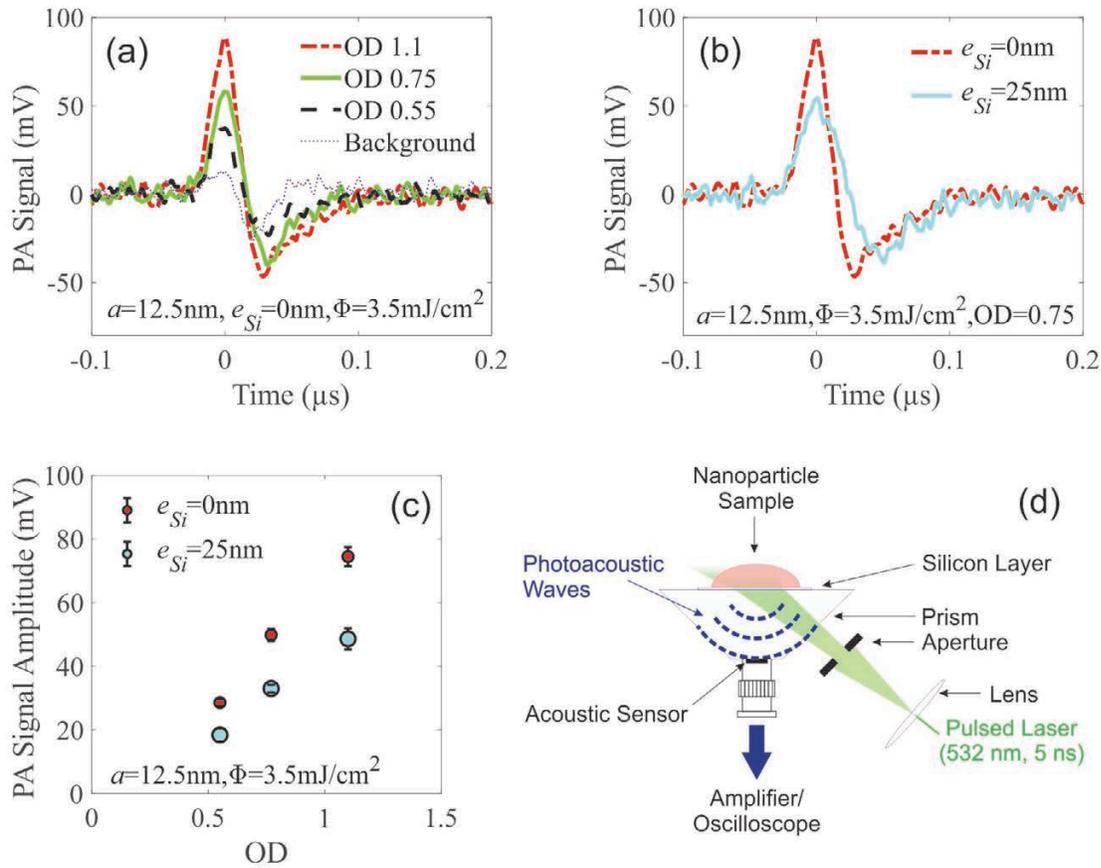

Figure 6. PA signal measurements of the in-house-synthesized gold nanosphere suspensions ($a = 12.5$ nm) using $\Phi = 3.5$ mJ/cm². (a) PA signals for uncoated gold nanosphere suspensions ($e_{Si} = 0$ nm) for three different OD representing different concentrations; the background PA signal with a non-absorbing sample on the sensor is also shown. (b) PA signals compared for samples at OD 1.1 with and without silica coating ($e_{Si} = 0$ and 25 nm). (c) The background-subtracted PA signal amplitude for the $a = 12.5$ nm, $e_{Si} = 0$ and 25 nm suspensions as a function of OD; the error bars represent the standard deviation of repeated measurements. (d) Schematic of the PA experimental setup.

The PA signal amplitude as a function of $\Phi$ from the uncoated and silica-coated nanosphere suspensions of identical OD is shown in Figure 7. The data from the in-house-synthesized particles ($a = 12.5$ nm, $e_{Si} = 0, 15,$ and 25 nm) are shown in Figure 7a, with all suspensions diluted to an OD of 0.75 (as shown in Figure 5b). The PA signal amplitude from each particle suspension is linear with $\Phi$ within the uncertainty limits, and the PA signal amplitude is again consistently lower for larger $e_{Si}$ when excited with the same fluence. Figure 7b shows the data for the commercially available nanosphere suspensions ($a = 5$ nm, $e_{Si} = 0$ and 17 nm), where each suspension was diluted to an identical OD of 0.87 (Supporting Information, Figure S1b). For the commercially available particle suspensions, the same trends appear as for the in-house-synthesized nanosphere suspensions, namely that the presence of a silica coating causes the PA signal amplitude to drop in comparison to the PA signal from the uncoated particle suspension. Similar trends were also found for both the in-house-synthesized particles and the purchased particles for additional data taken from suspensions diluted to a different OD (indicating a different concentration), as shown in the Supporting Information, Figure S3.

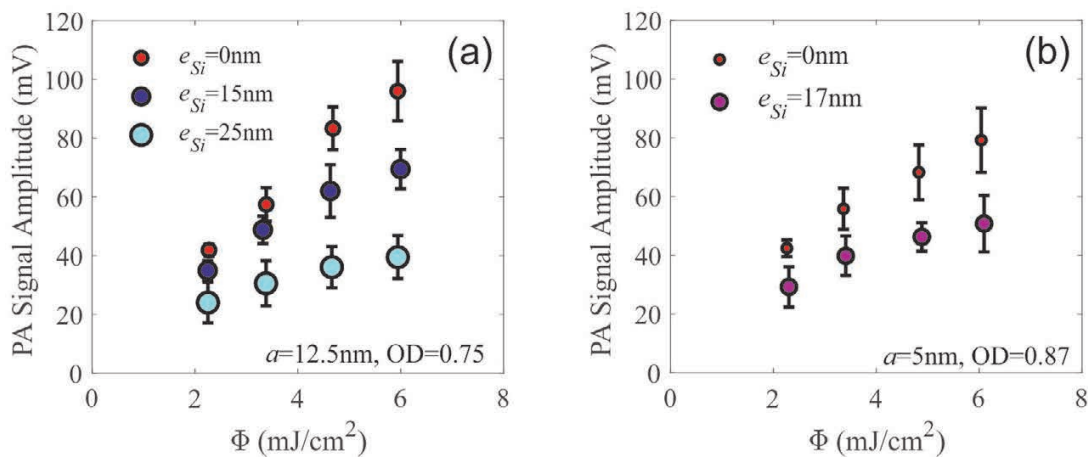

Figure 7. PA signal amplitude as a function of fluence from (a) the in-house-synthesized nanosphere suspensions $a = 12.5$ nm of OD 0.75 with $e_{Si} = 0$, 15, and 25 nm, and (b) the commercially available nanosphere suspensions $a = 5$ nm of OD 0.87 with $e_{Si} = 0$ and 17 nm.

**Comparison of Theoretical and Experimental Results**

As a main result of both our theoretical predictions and experimental measurements, the effect of a silica coating on a gold nanosphere is to reduce the efficiency of the PA generation. Figure 8a shows the measured PA signal amplitude from the silica-coated nanosphere suspensions for each silica-coating thickness ($e_{Si} = 15, 17$ and 25 nm), normalized by the signal amplitude from the respective uncoated nanosphere suspension ($e_{Si} = 0$ nm), for both the in-house-synthesized and commercially available nanosphere suspensions. The results shown are averaged over all the measurements with different Φ and OD, including five captured data traces for each of the measurement repetitions at each Φ and OD combination. The error bars represent the standard deviation of the normalized PA signal amplitude. The data is shown in comparison with the corresponding prediction from the theoretical model that was shown in Figure 4b. The trend seen from the experimental data are in qualitative agreement with the theoretical predictions, showing a decreasing PA signal amplitude with increasing $e_{Si}$. Within the experimental uncertainty limits of the measured PA signal amplitude as a function of $e_{Si}$ shown in Figure 8a, there is no distinguishable influence of changing the core gold particle diameter on the relative amount of signal reduction. This is consistent with the predictions of our theoretical model.

Quantitative deviations of the experimental data from the model could be accounted for by the fact that the model assumed the properties of the silica coating to be those of dense fused silica. The thermal conductivity of silica depends on its porosity, with a lower thermal conductivity for more porous silica. For example, Coquil et al.[35] measured thermal conductivity values of porous silica as low as 0.18 W m$^{-1}$ K$^{-1}$, a value 13% of the thermal conductivity of fused silica, for a porosity of 46%. Silica coatings synthesized using TEOS are accepted as being mesoporous.[9, 36-38] If the silica of our coated nanospheres was sufficiently porous, the model prediction of the normalized PA signal amplitude would be lower, in closer quantitative agreement with our data. This is illustrated in Figure 8b, which shows the model results computed with the thermal conductivity of the silica coating at 50% and 20% of its value for fused silica. Due to the lack of literature on reasonable values for the porosity of the silica coating synthesized using the Stöber method, we are unable to predict the precise magnitude of the effect of porosity.

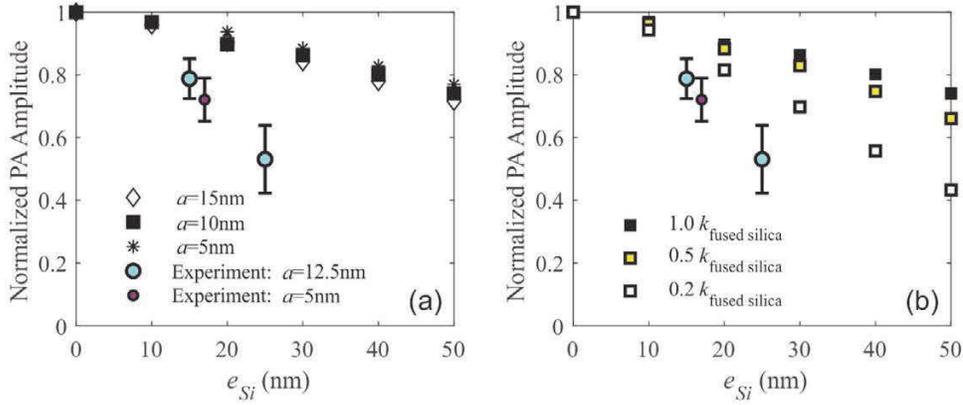

Figure 8. PA signal amplitude from the silica-coated gold nanospheres normalized by that from the respective uncoated gold nanospheres, as a function of $e_{Si}$; the error bars represent the standard deviation of the ratio computed with all experimental combinations of Φ and OD. Shown is the comparison to the theoretical prediction with (a) different core gold nanosphere sizes ($a = 5$ to $20$ nm), and (b) varying silica thermal conductivity ($k_s$) as a function of that of the thermal conductivity of fused silica ($k_{fused\ silica}$) for $a = 10$ nm.

It is also important to note that the model describes the PA signal for a single particle in water, while the experiments are for an aqueous suspension of an ensemble of particles. Also, there is a finite acoustic detector bandwidth in the experiments, and also a finite geometry of the detection surface area which could contribute to quantitative discrepancies between the experiment and the model, whereas the simulation results report the photoacoustic pressure at a single point in space without and frequency filtering. Nevertheless, it is clear from both our experimental data and theoretical model that the presence of a silica coating on a gold nanosphere causes a reduction in the PA signal amplitude, and that the signal amplitude decreases with increasing $e_{Si}$.

**Comparison to Previous Studies**

The recent simulation study from Kumar *et al.*[24] uses the commercial COMSOL software to perform finite-element-modeling-based simulations of the PA response to a 5 ns excitation pulse incident on core-shell particles, including silica-coated gold nanospheres. The trend of our temperature simulations, shown in Figure 3d, is in agreement with their results, showing that increasing $e_{Si}$ increases the temperature at a location in water away from the nanosphere center, confirming improvement in heat transfer with a silica coating. Their simulated PA signal results, however, contradict our findings, as they show that a silica coating provides an enhanced PA signal amplitude, with $e_{Si} = 20$ nm providing the best enhancement. Their explanation for the increased PA signal amplitude with silica coating was the reduction of the effect of Kapitza resistance. From the manuscript of Kumar *et al.*, it is unclear whether and how they simulated the Kapitza resistance at the gold-water interface, thus we cannot critically evaluate their simulation results compared to ours. Our theoretical results show that the expected Kapitza resistance does not contribute an observable effect on the PA signal from nanosecond excitation pulses. This prediction is in agreement with our previous work,[25] where the PA signal amplitude from suspensions of uncoated gold nanospheres was found to be similar to that of a homogeneous absorbing solution of the same absorption properties, which experimentally suggests a negligible effect of the Kapitza resistance on the PA signal from uncoated gold nanospheres in water. This conclusion is also further supported by additional experimental results reported in the Supporting Information (Figure S3), where the current measured PA signal amplitudes are similarly compared with those from a homogeneous absorbing aqueous dye of identical absorption properties. In addition to the insufficient description of the Kapitza resistance simulation from Kumar *et al.*, their simulated PA signals are also presented with a very coarse time step (approximately 1 ns). For the speed of sound and the apparent grid spacing of their reported simulation, this time step is typically not compatible with a stable converging solution (as a comparison, our study uses a time step on the order of tenths of a picosecond, chosen through an appropriate stability condition as discussed in the Supporting Information). This further casts doubts on the validity of their conclusions on the effect of silica coating on the PA signal generation.

A few previous experimental works have shown that PA signal amplitudes from gold nanospheres and nanorods coated with silica are enhanced.[5, 15, 16] Specifically, one experimental work on silica-coated nanosphere suspensions from Chen *et al.*[16] concluded that a silica coating enhanced the PA signal amplitude by up to a factor of three. The authors experimentally demonstrated such signal enhancement using gold nanospheres of $a = 13$ nm with $e_{Si} = 38$ nm. They found an enhanced PA signal amplitude for $e_{Si} \geq 18$ nm when the particle suspensions were illuminated by pulsed laser excitation from an Nd:YAG-pumped optical parametric oscillator around 530 nm (tuned to the peak resonance of the particles' extinction spectra) with $\tau_p = 5$ ns and $\Phi = 5$ mJ/cm². The authors hypothesized that the enhancement of their measured PA signal was caused by an increased heat transfer rate from the gold particle to the water due to the silica coating, which reduces the thermal resistance between an uncoated gold nanosphere and water. As mentioned in the preceding discussion, our theoretical model predicts that the Kapitza resistance at the gold-water interface has a negligible effect on the PA signal. In the case for an uncoated gold particle where Kapitza resistance at the particle-water interface is negligible, it has been mentioned by Chen and coworkers in an earlier publication[15] that "adding a shell of any material with finite heat conductance and capacity will only broaden a heat pulse and deteriorate the PA signal." Our own theoretical and experimental results are in agreement with this statement from Chen and coworkers, however, we reach the opposite conclusion regarding their experimentally observed effect of a silica coating on the PA generation from a gold nanosphere.[16] This therefore suggests that their experiments may have involved some additional physical phenomena, beyond those included a thermal diffusion and thermoelastic expansion. The presented data from Chen *et al.* only examine the PA signal at a single OD 1.5 and $\Phi = 5$ mJ/cm², and no comparison of their results with PA signals in a homogeneous absorbing solution were shown. As a result, it is difficult to speculate on what additional phenomena could be present in their experiments without knowing the absolute value of their measured PA signals in comparison to the homogeneous absorber case, and whether any nonlinear effects would have been observed from their samples.

In addition to literature on silica coating of gold nanospheres, several studies have also focused on silica coating on gold nanorods,[5, 6, 13-15] which are useful in biomedical imaging because the localized surface plasmon resonance for gold nanorods can be tuned to the infrared spectrum where a deep-tissue biological imaging window exists.[12] Moreover, silica coating is especially important for gold nanorods because the coating prevents deformation and reshaping of the nanorods.[7-10] Both Jokerst et al.[5] and Chen et al.[15] found enhanced PA signal amplitudes from silica-coated gold nanorod suspensions in comparison with uncoated gold nanorod suspensions, which was attributed to a reduction in the Kapitza resistance between the gold nanorod and surrounding medium due to the presence of the silica coating.[15] While our model shows that in the case of gold nanospheres, a reduction in the thermal resistance due to silica coating would not enhance the PA signal, our conclusion may not be directly applicable to silica-coated nanorods for two main reasons. First, the volume occupied by a coating scales roughly with $a^2$ for nanorods with a long aspect ratio, as compared with $a^3$ for nanospheres, meaning the influence of the finite volume occupied by the coating may be different. Second, uncoated gold nanorods are typically stabilized with a cetyl trimethylammonium bromide (CTAB) layer, which has been claimed to provide a higher thermal resistance compared to the short citrate ligands on uncoated gold nanospheres.[9]

Our results indicate that silica coating alone is not expected to lead to an enhanced PA signal in suspensions of coated gold nanospheres as compared to uncoated gold nanospheres. To achieve higher PA signal from nanoparticles through application of a coating, it could be advantageous to use a thin coating with high thermal conductance, such as a reduced graphene oxide coating as proposed by Moon *et al.*[39] Alternatively, a thermally confined shell that is more prone to thermal expansion, such as a polydimethylsiloxane coating as proposed by Shi *et al.,*[40] could be applied. Nevertheless, our findings are still important for imaging gold nanospheres used in biomedical application that may be coated with silica for other reasons, including increased colloidal stability and resistance to melting, low toxicity, improved functionalization possibilities, and increased biological uptake.[1-11] Whether our result can be extrapolated to silica coatings on other nanoparticle shapes that have found increasing interest in biomedical imaging, such as gold nanorods, remains a subject for further investigation.

## Conclusions

We developed a comprehensive model describing the PA signal generation from nanosecond-pulsed excitation of silica-coated gold nanospheres in water, and collected experimental data that supports the model predictions. The results from our theoretical and experimental studies both indicate that although the addition of a silica coating does enhance the heat transfer from the particle to its environment, the silica coating reduces the PA signal amplitude in comparison to the signal amplitude generated from a suspension of uncoated gold nanospheres with the same absorption cross section. The signal amplitude reduction increases as the silica coating thickness increases, and this was observed with both in-house-synthesized silica-coated gold nanosphere suspensions and commercially available silica-coated gold nanosphere suspensions. Our theoretical and experimental results contradict the common assertion that silica coating should increase the PA signal amplitude, and suggest that earlier experimental results showing such an increase cannot be explained through reduction of Kapitza resistance alone. Additional modeling needs to be done to determine if similar results can be expected for silica coating of gold nanorods and other nanoparticle shapes of interest in biomedical imaging. Although we did not find signal enhancement due to silica coating, our findings are still valuable for the biomedical PA imaging community because silica-coated gold nanoparticles have additional advantages to be used in biomedical imaging, and our results are critical in facilitating the interpretation of PA signals when these particles are applied in biomedical studies.

## Experimental Methods

**Nanosphere Synthesis and Characterization**

For the in-house-synthesized set of uncoated and silica-coated gold nanosphere suspensions of $a = 12.5$ nm, chloroauric acid (HAuCl4), sodium citrate, mPEG-thiol (average $M_n$ 6000), ammonia, tetraethyl orthosilicate (TEOS) and isopropanol were purchased from Aldrich. Citrate-stabilized gold nanospheres of $a = 12.5$ nm were synthesized according to the procedure described in Wang et al.[41] Briefly, the glassware and magnetic stirrer used for the synthesis was cleaned thoroughly with aqua regia (HNO$_3$-HCl, 1:3, v/v) and then washed with distilled water. 50 µL of HAuCl4(aq) (0.01%) was boiled in an Erlenmeyer flask, and 0.75 µL sodium citrate was added under vigorous stirring. The color of the solution changed from yellow to black to red, after which the solution remained under continuous boiling and stirring for an additional 15 minutes. The resulting gold nanosphere suspension was cooled and stored at 4 °C, forming the in-house-synthesized uncoated gold nanosphere suspension of $a = 12.5$ nm, confirmed by SEM images (Figure 5). PEGylated gold nanospheres were then produced by exchanging citrate with mPEG-thiol such that silica could be coated directed on the PEGylated gold nanospheres using a modified Stöber method.[32, 33] Briefly, the citrate-stabilized uncoated gold nanosphere suspension was centrifuged (Eppendorf 5804 R, 12000 rpm for 15 minutes) before the supernatant was removed and the particles were redispersed in distilled water. Then 3 mL of mPEG-thiol (0.2 mM) was added to this suspension under vigorous stirring before the mixture was sonicated for 5 minutes in an ultrasound bath and then allowed to react overnight. The excess mPEG-thiol was then removed through centrifugal filtration (12000 rpm for 15 minutes) and the PEGylated gold nanospheres were redispersed in distilled water. To prepare the silica-coated gold nanoparticles, 1.2 mL of the PEGylated gold nanosphere suspension was added to 1.8 mL of isopropanol under vigorous stirring. A freshly prepared solution of ammonia (3.8% in isopropanol) was added until the solution reached a pH value of 11 (100 µL of the ammonia sufficed in our case). A freshly prepared TEOS solution (100 mM) was then added to the solution, which remained under vigorous stirring for three hours. For the $e_{Si} = 15$ nm silica coating thickness, 0.4 mL of the TEOS solution was used. For the $e_{Si} = 25$ nm silica coating thickness, 0.6 mL of the TEOS solution was used. After the reaction, the suspension was centrifuged (12000 rpm for 15 minutes) before the supernatant was removed and the particles were redispersed in 3 mL of distilled water. Each suspension was sonicated 5 minutes in an ultrasound bath after the original synthesis, and again for one minute before each photoacoustic experiment. All PA experiments and the imaging of the nanosphere suspensions were carried out within 60 days of the silica-coated particle synthesis.

The commercially available set of uncoated and silica-coated gold nanosphere suspensions were obtained from Aldrich (Silica-coated: 747564, 10 nm diameter, silica coated, OD 1, dispersion in $H_2O$, Uncoated: 741957, 10 nm diameter, OD 1, stabilized suspension in citrate buffer). The as-purchased suspensions were diluted in distilled water to the desired OD. Each suspension was sonicated for one minute before each photoacoustic experiment to precluded particle aggregation in our suspensions.

The core gold nanosphere diameter and silica coating thickness for each of the particle suspensions were confirmed through scanning electron microscopy imaging (GeminiSEM, Carl Zeiss, Jena, Germany). The OD of the suspensions were measured in a UV/Vis Spectrometer (Specord 250 Plus, Analytik Jena AG, Jena, Germany) using a disposable polymethylmethacrylat cuvette with 10 mm optical path length. The overall optical extinction and absorption in the silica-coated particle suspensions was measured in an integrating sphere, following the procedure in our previous publication.[25]

**Photoacoustic Signal Measurement**

PA signals were measured in a backwards mode PA sensor with pulsed nanosecond laser light excitation at 532 nm, similar to our previous work.[25] Briefly, the sensor head is based on a 5-mm diameter circular 25-μm-thick piezoelectric poly(vinylidene) fluoride (PVDF) foil coupled to the bottom of a transparent glass prism with conductive epoxy. The PVDF foil is integrated into a BNC socket of which the output was amplified and captured by a digitizing oscilloscope (Tektronix TDS 620A). Above the prism directly across from the PVDF foil is a 1-mm-thick clear silicon layer acoustically coupled with water to hold a free-sitting 300 μL drop of the sample. The PA excitation at 532 nm was achieved using a Q-Switched, frequency-doubled Nd:YAG laser (Surelite 10-I, Continuum, Santa Clara, CA) with $\tau_p = 5$ nm (fwhm) and a repetition rate of 10 Hz. The laser beam was shaped through a series of apertures and lenses, and entered the base of the prism, on the same side as the acoustic detection. A homogeneous area of 4 mm in diameter was illuminated at the sample surface, which was confirmed using a laser beam profiler (LaserCAM, Coherent). The laser energy was controlled by adjusting the Q-Switch delay of the laser to obtain Φ at the sample surface between 2 and 6 mJ/cm². A portion of the beam upstream of the PA sensor was diverted with a beam splitter into a pyroelectric energy probe (Laser Probe

RjP-735, Utica, NY) and monitored through a universal radiometer (Laser Probe Rm-6600, Utica, NY) connected to the oscilloscope.

The PA signal traces and laser energy data from the oscilloscope were acquired using an in-house developed LabVIEW program that captured 50-pulse averages for each saved measurement repetition. All PA signal measurements were repeated at least three times with new nanoparticle samples, and five PA signal traces were saved for each measurement repetition. Before the PA signal from each new nanoparticle suspension was measured, an Eppendorf tube containing the suspension was placed in an ultrasound bath for up to one minute to break up any potential aggregated particle clusters.

## Acknowledgements

This project has received funding from the European Research Council (ERC) under the European Union's Horizon 2020 research and innovation program (grant 681514, COHERENCE). G. Pang acknowledges funding provided by an Alexander von Humboldt Fellowship for Postdoctoral Researchers and support from host R. Nießner. X. Wang and S. Mahler helped with the nanoparticle synthesis. C. Sternkopf performed the SEM imaging of the nanoparticles. Feedback from A. Desjardins was helpful for developments in the project.

## Supporting Information Available

A Supporting Information file provides details on the theoretical models including impulse response solutions, discretized equations, conversion criteria, and solution procedures, and provides supplementary figures showing characterization of the commercially available nanosphere samples, extinction and absorption measurements using an integrating sphere, and additional PA signal amplitude data. A Matlab code that computes the solutions of the theoretical models is available on request.

# SUPPORTING INFORMATION FOR:

Theoretical and experimental study of photoacoustic excitation of silica-coated gold nanospheres in water

*Genny A. Pang[1a], Florian Poisson[2], Jan Laufer[3], Christoph Haisch[1], Emmanuel Bossy[2b]*

[1] Chair for Analytical Chemistry and Institute of Hydrochemistry, Technische Universität München, Marchioninistraße 17, 81377, Munich, Germany

[2] Univ. Grenoble Alpes, CNRS, LIPhy, 38000 Grenoble, France

[3] Institut für Physik, Martin-Luther-Universität Halle-Wittenberg, Halle (Saale), Germany

[a] E-Mail: genny.pang@tum.de

[b] E-Mail: emmanuel.bossy@univ-grenoble-alpes.fr

**Theory Details and Solution Procedure**

The generation of a photoacoustic wave from a single silica-coated gold nanosphere in water was modeled assuming spherical symmetry, with the center of the gold nanosphere as the center of symmetry. The conservation of energy within the gold nanoparticle and the thermal diffusion equation in water and silica are given by Equations (1) and (2) in the manuscript, respectively, and the boundary condition are given by Equations (3) and (4) in the manuscript. A semi-analytical solution of Equations (1) through (4) to determine the temperature field in time and space was obtained through an analogous approach as done by Prost et al.[21] for the problem without silica coating. The basic procedure is to use the Laplace Transform method for solving differential equations to obtain the Green's function $G(\vec{r}, t)$, which is the solution to an impulse (Dirac delta function) response. The temperature field as a response to a pulsed excitation is then solved through numerical convolution in Matlab of the thermal Green's function with a Gaussian source term, as given in Equation (S5).

$$T_i(r, t) = G_i(r, t) * \frac{2\sqrt{\ln(2)}}{\sqrt{\pi}} e^{-4\ln 2 \left(\frac{t}{\tau_p}\right)^2} \qquad (S1)$$

where $i = g$, $s$ or $w$ for gold, silica and water respectively. The temperature profile as a function of time and space was computed and sampled on a regular grid with spatial and temporal intervals matching those required for the finite-difference in time resolution of the thermoelastic problem.

The Green's functions solution obtained by using the Laplace Transform are obtained for two cases: **[A]** no silica coating and finite thermal resistance at the gold-water interface, and **[B]** silica coating and no thermal resistance at both the gold-silica and silica-water interfaces. The equations were written with the following dimensionless variables:

$$\hat{t} = \frac{t}{\tau_w}, \qquad \tau_i = \frac{a^2}{\alpha_i}, \qquad \alpha_i = \frac{k_i}{\rho_i c_{pi}}$$

$$\hat{T}(t) = \frac{T(t)-T_0}{T_{max}}, \text{ with } T_{max} = \frac{3\sigma_a \Phi_0}{4\pi a^3 \rho_g c_{pg}}$$

$$\hat{r} = \frac{r}{a}, \qquad \hat{R} = \frac{a+e_{Si}}{a}$$

$$\chi_{i-j} = \frac{3\rho_j c_{pj}}{\rho_i c_{pi}}, \qquad \gamma_{i-j} = \frac{k_i}{k_j}, \qquad B = \frac{a}{k_g R_{th\,g-w}}$$

For Case **[A]**, the solution to the thermal problem is given by:

$$\hat{G}_g(\hat{t}) = \frac{1}{\pi}\int_{-\infty}^{\infty} \mathbb{R}\left[\frac{1+B+iu}{Q_A(u)} iu e^{-u^2 \hat{t}}\right] du \tag{S6}$$

$$\hat{G}_w(\hat{r} > 1, \hat{t}) = \frac{1}{\hat{r}}\cdot\frac{1}{\pi}\int_{-\infty}^{\infty} \mathbb{R}\left[\frac{B}{Q_A(u)} e^{-iu(\hat{r}-1)} iu e^{-u^2 \hat{t}}\right] du \tag{S7}$$

with $Q_A(u) = -u^2(1 + B + iu) + \chi_{g-w} B(1 + iu)$

And for Case **[B]**, the solution to the thermal problem is given by:

$$\hat{G}_g(\hat{t}) = \frac{1}{\pi}\int_{-\infty}^{\infty} \mathbb{R}\left[\frac{D_+(u,1)}{Q_B(u)} iu e^{-u^2 \hat{t}}\right] du \tag{S8}$$

$$G_s(\hat{t}, 1 \leq \hat{r} \leq \hat{R}) = \frac{1}{\pi}\frac{1}{\hat{r}}\int_{-\infty}^{\infty} \mathbb{R}\left[\frac{D_+(u,\hat{r})}{Q_B(u)} iu e^{-u^2 t}\right] du \tag{S9}$$

$$\hat{G}_w(\hat{t}, \hat{r} \geq \hat{R}) = \frac{1}{\pi}\frac{1}{\hat{r}}\int_{-\infty}^{\infty} \mathbb{R}\left[\frac{\left(-2\gamma_{s-w}\hat{R}iu\sqrt{\frac{\tau_s}{\tau_w}}\right)}{Q_B(u)} e^{-iu(\hat{r}-\hat{R})} iu e^{-u^2 t}\right] du \tag{S10}$$

with the following definitions:

$$A_-(u) = -(1 + \hat{R}iu) + \gamma_{s-w}\left(1 - \hat{R}iu\sqrt{\frac{\tau_s}{\tau_w}}\right), \quad A_+(u) = (1 + \hat{R}iu) - \gamma_{s-w}\left(1 + \hat{R}iu\sqrt{\frac{\tau_s}{\tau_w}}\right)$$

$$D_+(u,\hat{r}) = A_+(u) e^{iu\sqrt{\frac{\tau_s}{\tau_w}}(1-\hat{R})} + A_-(u) e^{-iu\sqrt{\frac{\tau_s}{\tau_w}}(1-\hat{R})}$$

$$D_-(u,\hat{r}) = A_+(u)e^{iu\sqrt{\frac{\tau_s}{\tau_w}}(1-\hat{R})} - A_-(u)e^{-iu\sqrt{\frac{\tau_s}{\tau_w}}(1-\hat{R})}$$

$$Q_B(u) = \left(\frac{\tau_w}{\tau_s}\chi_{g-s} - u^2\right)D_+(u,1) - iu\sqrt{\frac{\tau_w}{\tau_s}}\chi_{g-s}D_-(u,1)$$

The integrals in Equations (S6) to (S10) were solved numerically in Matlab. The Green's function as a function of dimensioned variables r, and t is given

$$G_i(r,t) = \hat{G}_i(\hat{r},\hat{t}) \times T_{max} + T_0 \qquad (S11)$$

where $T_{max}$ is the temperature that would be reached by the particle without any cooling, and $T_0$ is the background temperature. The solution to a third case, no silica coating and no thermal resistance, derived analogous to Case **[A]** and **[B]** and given in Prost et al.[21], was also used in our simulations for comparison.

The governing equations describing thermoelastic expansion and propagation in the gold, silica, and water are described by Equations (5) to (7) in the manuscript. They were solved numerically with a finite-difference time-domain (FDTD) approach, with the same numerical scheme described in detail in Prost et al.[21] (see Appendix C of reference), with all thermal expansion coefficient $\beta$ taken as independent of temperature. The temporal step $\Delta t$ of the FDTD approach was chosen by the stability condition presented in Prost et al.[21] (Appendix C), given by

$$\Delta t = 0.99 \frac{\Delta r}{\sqrt{3}c_{Au}} \qquad (S12)$$

where $\Delta r$ is radial grid spacing and $c_{Au}$ is the speed of sound in gold, which is the highest speed of sound involved in the problem. For the current simulations, $\Delta r = 1$ nm was used in all cases, except for the simulations with finite Kapitza resistance where $\Delta r = 0.5$ nm was used.

**Supplementary Figures**

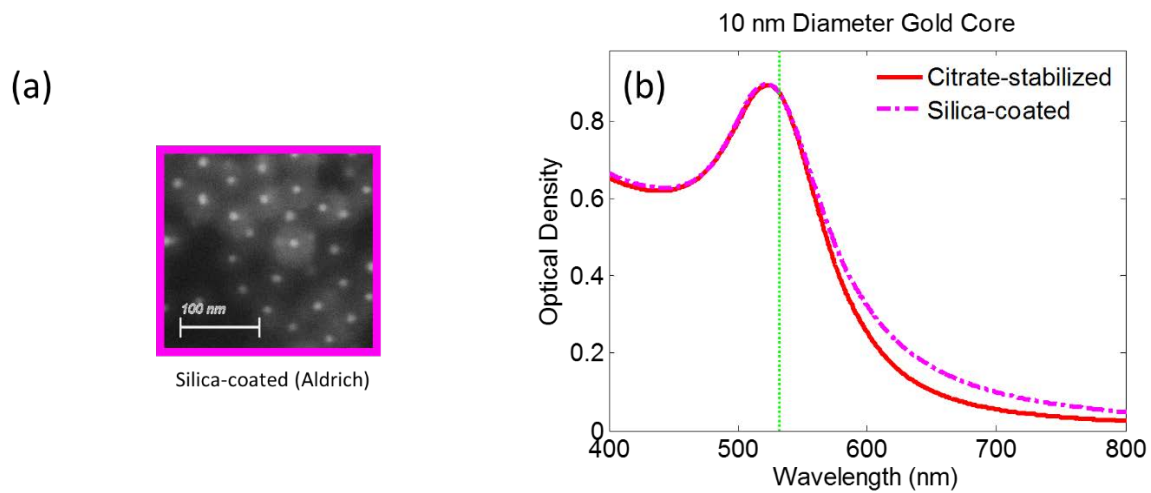

Figure S1. (a) SEM images of silica-coated gold nanoparticles purchased from Sigma Aldrich, (b) OD spectra of bare and silica-coated gold nanoparticles from Sigma Aldrich ($a = 5$ nm) measured in a 10 mm cuvette where both suspensions have an OD of 0.87 at the measurement wavelength of 532 nm (dotted green line).

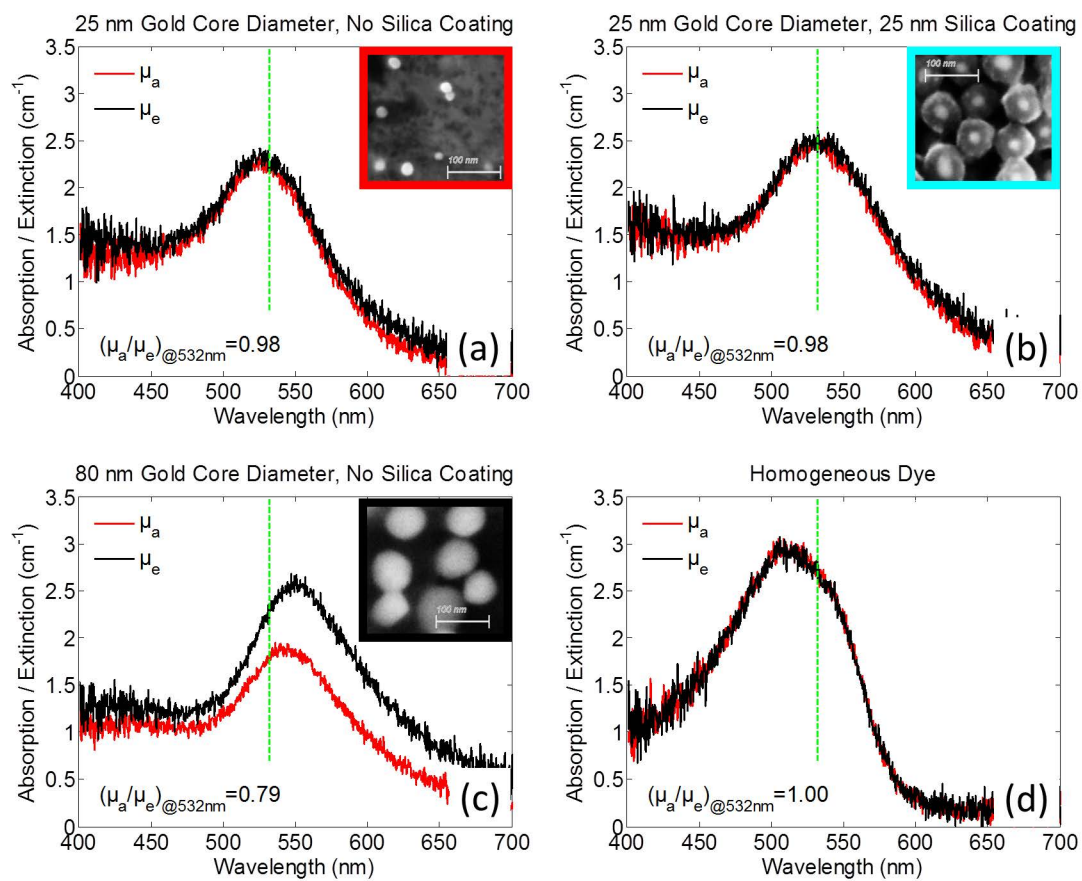

Figure S2. Measurements of optical extinction and absorption using an integrating sphere for (a) in-house-synthesized gold nanospheres with no silica coating ($a = 12.5$ nm, $e_{Si} = 0$ nm), (b) in-house-synthesized 25 nm gold core diameter nanospheres with silica coating ($a = 12.5$ nm, $e_{Si} = 25$ nm), (c) large gold nanospheres with no silica coating of similar total diameter as the silica-coated nanospheres ($a = 40$ nm, $e_{Si} = 0$ nm), and (d) an aqueous homogeneous absorbing dye solution (Sirius® Supra Red F4BL).

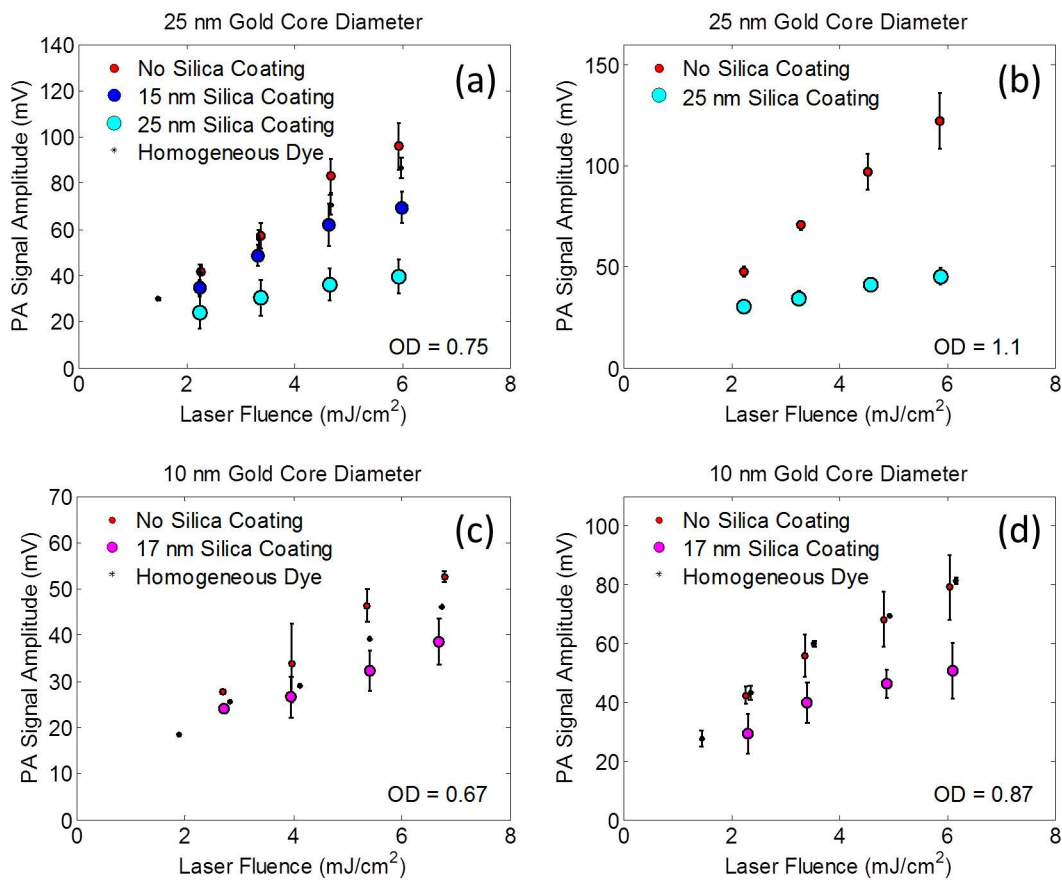

Figure S3. All PA signal amplitude data for (a) the in-house-synthesized silica-coated gold nanospheres ($a = 12.5$ nm) with OD 0.75 and (b) 1.1, and (b) the commercially available silica-coated gold nanospheres ($a = 5$ nm) with OD 0.67 and (d) 0.87. Also shown is the PA signal amplitude from an aqueous homogeneous absorbing dye solution of the same respective OD.

7